\documentclass[a4paper,11pt,reqno]{amsart}

\usepackage[english]{babel}
\usepackage[latin1]{inputenc}
\usepackage{a4wide}
\usepackage{amsfonts}
\usepackage{amsmath}
\usepackage{amssymb}
\usepackage{amsthm}
\usepackage{graphicx}
\usepackage{color}

\begin{document}

 \author{Gernot Bauer}
 \email{gernot.bauer@fh-muenster.de}
\address{Fachhochschule Münster, Bismarckstr. 11, 48565
Steinfurt, Germany.}
 
\author{Dirk - Andr\'e Deckert}
 \email{deckert@math.ucdavis.edu}
\address{Department of Mathematics,
University of California,
One Shields Avenue,
Davis, 95616, CA, USA}

 \author{Detlef D\"urr}
 \email{duerr@mathematik.uni-muenchen.de}
\address{Mathematisches Institut der LMU, Theresienstr. 39, 80333
M\"unchen, Germany.}

\author{G\"unter Hinrichs}
\email{hinrichs@math.lmu.de}
\address{Mathematisches Institut der LMU, Theresienstr. 39, 80333
M\"unchen, Germany.}

\title{On irreversibility and radiation in classical electrodynamics of point particles}

\begin{abstract}
The direct interaction theory of electromagnetism, also known  as Wheeler-Feynman electrodynamics, is often misinterpreted and found unappealing because of its reference to the absorber and,
more importantly, to the so-called absorber condition.
Here we remark that the absorber condition  is indeed questionable and presumably not relevant for the explanation of irreversible radiation phenomena in our universe.
What is relevant and deserves further scrutiny is the emergent effective description of a test particle in an environment.
 We therefore rephrase what we consider the relevant calculation by Wheeler and Feynman and comment on the status of the theory.\\

\end{abstract}\footnote{Dedicated to Herbert Spohn on the occasion of his 65th birthday. } 
\maketitle


\section{Introduction/outline}\label{sec:introduction}
It is well known that the coupled Maxwell-Lorentz equations for  interacting point particles are ill-defined because the field diverges at the location of its sources.
The so-called Wheeler-Feynman electrodynamics, which actually goes back to Schwarzschild and Fokker (\cite{Schwarzschild,Fokker}), represents a physically relevant
and mathematically well-defined as well as simple theory of relativistic interaction in which this problem is not just cured but absent from the beginning.
So far, it is the only theory of classical electrodynamics that has been shown to be capable of predicting radiation phenomena.
It is given by the following equations of motion
\begin{align}\label{Gleichung}
 m_i\ddot z_i^\mu(\tau_i) = e_i\sum_{j\in\{1,\dots,N\}\backslash {i}}\frac12\left[F_{j,+}+F_{j,-}\right]^{\mu\nu}(z_i(\tau_i))\dot z_{i,\nu}(\tau_i) \hspace{1cm} (i\in\{1,\dots,N\})
\end{align}
with
\begin{equation}\label{F}
 F_{j,\pm}^{\mu\nu}(z) := \frac{\partial A_{j,\pm}^\nu}{\partial z^\mu}(z) - \frac{\partial A_{j,\pm}^\mu}{\partial z^\nu}(z)\,,
\end{equation}
where
\begin{equation}\label{A}
 A_{j,\pm}^\mu(z) := \frac{e_j\dot z_j^\mu(\tau_{j,\pm}(z))}{(z-z_j(\tau_{j,\pm}(z)))_\nu \dot z_j^\nu(\tau_{j,\pm}(z))}
\end{equation}
and the \emph{advanced and retarded times} $\tau_{j,\pm}(z)$ are defined implicitly by
\begin{equation*}
 (z-z_j(\tau_{j,\pm}))_\mu (z-z_j(\tau_{j,\pm}))^\mu = 0 \,\,, z_j^0(\tau_{j,+})>z^0, z_j^0(\tau_{j,-})<z^0\,.
\end{equation*}
This theory describes  a total number of $N$ interacting point particles with masses $m_i$, charges $e_i$ and world lines $z_i^\mu$ in Minkowski space parametrized by their proper times $\tau_i$. Here, we use the Einstein notation $x_\mu y^\mu:=\sum_{\mu=0}^3 x_\mu y^\mu$.

Wheeler-Feynman electrodynamics contains no fields. Charges interact directly whenever their corresponding space-time points have zero Minkowski distance.
A formal way to present the theory is via an action principle from which the above equations arise as Euler-Lagrange equations
minimizing the following action by variation over the world lines
$z_i^\mu(\lambda_i)$:

\begin{equation}\label{Wirkung}
S=\sum_i\left[ -m_i c\int\left(\dot z_i^\mu\dot z_{i\mu}\right)^{\frac{1}{2}}d\lambda_i
-\sum_{j>i}\frac{e_i e_j}{c}\int\int\delta\left((z_i-z_j)^2\right)\dot z_i^\mu\dot z_{j\mu}
d\lambda_i d\lambda_j\right].
\end{equation}

Note that self-interactions are explicitly excluded as the summation in \eqref{Gleichung} or \eqref{Wirkung} runs over $i\neq j$.

To connect this theory to radiation phenomena, which are well described on a phenomenological level by Maxwell-Lorentz electrodynamics, several issues need to be addressed:
\begin{enumerate}
\item[(I1)] Radiation damping occurs instantaneously with the acceleration of the source, is completely determined by its motion and independent of the environment.
How can a theory that explicitly excludes self-interaction account for these features?
\item[(I2)] Radiation seems to be directed from the past to the future, which in Maxwell-Lorentz electrodynamics is 
introduced by a restriction of the theory  to retarded potentials. How is this electromagnetic arrow of time explained in 
Wheeler-Feynman electrodynamics, which is completely time-symmetric?
\item[(I3)] Radiation is accompanied by loss of energy and causes radiation friction of the radiating particle. This friction is 
historically computed from a  self-interaction term by a formal method of mass renormalization \cite{Dirac38}. It is believed to be described by the so called Lorentz-Dirac equation, i.e.,
\begin{equation}\label{LD}
m\ddot x^\mu(\tau) = \frac{2e^2}{3c^3} \left(\dddot x^\mu(\tau) + \ddot x_\nu(\tau)\ddot x^\nu(\tau) \dot x^\mu(\tau)  \right)\,,
\end{equation}
which leads to an energy loss rate given by the experimentally well-verified Larmor formula
\begin{equation*}
P=-\frac{2e^2}{3c^3} \ddot x_\mu \ddot x^\mu \,.
\end{equation*}
The Lorentz-Dirac equation, however, has the very unpleasant feature that it involves a third derivative and gives rise to unphysical solutions (known as runaway solutions \cite{Jackson},
for a mathematical analysis of the equation see \cite{Spohn}), which have to be ruled out by an extra condition.
Does this equation emerge in a thermodynamic sense from Wheeler-Feynman electrodynamics as an effective equation?
-- If yes and if Wheeler-Feynman solutions show no runaway behaviour \cite{Bauer}, then the Wheeler-Feynman law \eqref{Gleichung} already rules out such unphysical solutions.
\end{enumerate}

The issues (I2) and (I3) are of course intertwined. Both are connected to irreversibility of motion,
and since Wheeler-Feynman electrodynamics is time-symmetric one faces the problem of explaining irreversibility in a time-reversible theory. 
In classical mechanics, however, one already knows how to do this - an explanation amounts to the second law of thermodynamics, i.e., the increase of entropy.
There we refer to the phase space of the particles of the universe, the points of which uniquely determine the evolution of the universe,
and the second law is explained by a particular low entropy ``initial state'' of the universe, a state with very low phase space volume. 
The idea being that the trajectories  will explore in time  ever larger regions of phase space, thus, increasing entropy, which is directly related to phase space volume occupied by the macrocopic
state of the universe.

If one wants to argue similarly in Wheeler-Feynman electrodynamics one immediately encounters the problem that no phase space description is known.
In fact, its mathematical analysis concerning existence and uniqueness of solutions is extremely hard. It is of course also hard for classical interacting theories like gravitating masses.
However, Wheeler-Feynman electrodynamics is not given by ordinary differential equations.
The presence of delayed terms have yet prevented the development of any general existence theory of solutions (though special solutions are known: \cite{schild_electromagnetic_1963}, \cite{Bauer}, \cite{martinez_2003}, \cite{DirkNicola}).
Especially, it is unknown what data is necessary to uniquely characterize the trajectories.
E.g., it might be the asymptotic behaviour (\cite{Bauer}), trajectory strips
(as has been shown in \cite{DirkNicola} in a toy model) or positions and velocities at space-like separation (\cite{Driver}).
Therefore, a definition  of entropy, based on a measure of typicality, analogous the phase space volume in classical mechanics based on Liouville's theorem is out of sight at the moment.

Further, we  stress that the whole  problem originates in relativity  rather than in the ``strangeness'' of the equations:
Introducing  fields as dynamical degrees of freedom, which mediate the interaction between the particles, reinstates a ``Markovian'' phase space description,
however, at the price of infinite self-interaction. A noteworthy exception is Maxwell-Lorentz theory without self-interaction (ML-SI, see \cite{ML-SI}) of which the WF trajectories are a subclass of solutions, see \cite{Dirk2}.

We can nevertheless discuss the issue of irreversibility in a handwaving manner and point towards a sensible solution.
Since this theory does not contain fields, they can only emerge as phenomenological descriptions on a macroscopic level.
So all there is are particles which move.
It is therefore tempting to think that the low entropy state of the distribution of matter used in classical physics should also be responsible for the arrow of time in Wheeler-Feynman electrodynamics. 

For our analysis of irreversible behavior it will be convenient to read the Wheeler-Feynman equations \eqref{Gleichung} in the language of fields even though fields are not part of the theory.
Since \eqref{F}-\eqref{A} coincides with the formula for the Liénard-Wiechert field generated by the particle with world line $z_j$,
we attribute such a field to every particle, so that every particle moves under the force generated by the fields of all other particles.
We hence  rewrite the equations in three-dimensional notation,
referring to the electric and magnetic Liénard-Wiechert fields associated to each particle $j$ by $ E_{j,\pm}$ and $ B_{j,\pm}$:
\begin{equation}\label{LW3d}
\begin{split}
&m_i\frac{\text d}{\text dt}\left(\frac{{ \dot x_i}(t)}{\sqrt{1-\frac{\|\dot x_i(t)\|^2}{c^2}}}\right) =
\frac{e_i}2 \sum_{j\neq i}\left( ({ E_{j,-}} + { E_{j,+}})(t,x_i(t)) +\frac{{ \dot x_i}(t)}{c} \times ({ B_{j,-}} + { B_{j,+}})(t,x_i(t)) \right) \\
&{ E_{j,\pm}}(t,x) := e_j\left[\frac{({ n_j}\pm \frac{ \dot x_j}c)(1-\frac{\|\dot x_j\|^2}{c^2})}{\|x-x_j\|^2(1\pm n_j\cdot \frac{\dot x_j}{c})^3} + \frac{\frac{ n_j}c\times[({ n_j}\pm\frac{ \dot x_j}c)\times\frac{ \ddot x_j}c]}{\|x-x_j\|(1\pm n_j\cdot\frac{\dot x_j}c)^3} \right] (\tau_{j,\pm}(t,x))\\
&{ B_{j,\pm}}(t,x) := \mp \frac{{ n_j}(\tau_{j,\pm}(t,x))}c\times { E_{j,\pm}}(t,x) \\
&{ n_j}:=\frac{{ x-x_j}}{\|x-x_j\|}
\end{split}
\end{equation}


Note that $ E_{j,\pm}$ and $ B_{j,\pm}$ can be decomposed naturally into a \emph{short-ranged field} given by the summand proportional to $\frac1{\|x-x_j\|^2}$ and a
\emph{long-ranged field} given by the summand proportional to $\frac1{\|x-x_j\|}$.


Establishing irreversible behavior is part of the more general question of how one can effectively describe test particles or subsystems
without detailed knowledge about the state of the whole universe. In a Newtonian universe for example we would hold the global non-equilibrium responsible for special situations like when
``a glass is put on a table and then pushed off its edge''. To follow the shattering of the glass, the pieces of which will typically not assemble themselves to form the glass again,
it is assumed that the ``rest of the universe'' does not interfere anymore with this process.
In this respect, the earliest approach in Newtonian gravity was to assume that the rest of the universe  approximately forms a continuous uniform mass shell of a certain thickness around the subsystem.
Then the forces exerted by the shell from any solid angle element $\text d\alpha$ on a particle surrounded by the shell
have all the same strength irrespectively of the distance $r$ between the particle and the shell segment-- a single shell particle exerts a force proportional to $\frac1{r^2}$ and the particle density is proportional to the area $4\pi\text d\Omega r^2$ of the shell segment,
so the total force is proportional to $\text d\Omega$.
Thus, forces from opposite directions cancel, the shell exerts no net force and the particles in the subsystem only feel the forces they exert on each other
(as long as they do not deform the shell too much).
On the one hand, the presence of ``shells'' and ``subsystems'' is a global non-equilibrium suggests a distribution of matter that can create ``special situations''
like a glass being put on a table and then pushed off its edge,
but is otherwise not interfering too much with the evolution of the subsystem when ``left to its own''.

Can a similar reasoning be applied in electrodynamics?
In analogy to the argument above, we would also hold the low entropy state of the universe  responsible for creating special motions of subsystems like, e.g.,
particle accelerators or, on a larger scale, solar winds. The special motion of particles in a subsystem would then initiate a manifold of motions of particles in the rest of the universe,
a process that similarly to the shattering of the glass, should typically occur in this ``forward'' direction only, where we assume again that the rest of the universe does not interfere anymore with it.
In order to give a similar argument as for Newtonian gravity also for the Wheeler-Feynman electrodynamics, we need to deal with an  additional difficulty as according to \eqref{LW3d},
the forces exerted by particles depend not only on their position, but also on their state of motion (velocity and acceleration).
Assuming that either all particles are at rest or the motion in the shell is such that one finds the same spherically symmetric velocity distribution in every small segment, 
one readily finds that the short-ranged part of the force on a particle on the inside cancels as it does in Newtonian gravity:
Call $y$ the degree of freedom corresponding to the velocity distribution at a point in the shell specified by the angle coordinate $\Omega$.
Then $\int \frac{1-\frac{\|\dot x\|^2}{c^2}}{\left(1\pm n\cdot\frac{\dot x}{c}\right)^3} \text dy=: C$ does not depend on $\Omega$, so
$$\int \frac{{ n} \left(1-\frac{\|\dot x\|^2}{c^2}\right)}{r^2\left(1\pm n\cdot\frac{\dot x}{c}\right)^3}\text dyr^2\text d\Omega = \int { n}C\text d\Omega = 0$$
as in Newtonian gravity.
Moreover, any term $\frac{\pm\frac{\dot { x}}{c}\left(1-\frac{\|\dot x\|^2}{c^2}\right)}{r^2\left(1\pm n\cdot \frac{\dot x}{c}\right)^3}r^2\text d\Omega$ cancels with
$\frac{\pm\frac{-\dot { x}}{c}\left(1-\frac{\|-\dot x\|^2}{c^2}\right)}{r^2\left(1\pm (-n)\cdot \frac{-\dot x}{c}\right)^3}r^2\text d\Omega$, so that
$$\int \frac{\pm\frac{\dot { x}}{c}\left(1-\frac{\|\dot x\|^2}{c^2}\right)}{r^2\left(1\pm n\cdot \frac{\dot x}{c}\right)^3}r^2\text d\Omega\text dy = 0\,.$$
Finally, one finds
$$\mp\frac{ n}{c} \times \frac{({ n}\pm \frac{ \dot x}c)(1-\frac{\|\dot x\|^2}{c^2})}{r^2(1\pm n\cdot \frac{\dot x}{c})^3} = \frac{{ n}\times{ \dot x}\left(1-\frac{\|\dot x\|^2}{c^2}\right)}{c^2r^2\left(1\pm n\cdot\frac{\dot x}{c}\right)^3} $$
and any retarded term (the ones with minus sign) cancels with a corresponding advanced term in which $ \dot x$ is replaced by $-\dot x$.

Under the additional, but more demanding assumption that the accelerations are distributed symmetrically and independently of the velocities in every small segment, also the far-field term vanishes, e.g.
\begin{equation*}
 \int \frac{\frac{ n}c\times[({ n}\pm\frac{ \dot x}c)\times\int\frac{ a}c\text da]}{r(1\pm n\cdot\frac{\dot x}c)^3}\text dyr^2\text d\Omega =0\,.
\end{equation*}
In conclusion, the force on the source vanishes.

The employed assumptions seem reasonable as long as the interaction between shell and test particles is negligible,
but what happens if one  allows for the special situation of atypically large accelerations of the test particles:
These lead to components of motion in the shell which are no longer spherically symmetric, but directed according to the axes of motion of the test particles.
Therefore, if a test particle experiences a kick, the shell is expected to exert a net reactive force on it.
Since according to \eqref{Gleichung}, influences are directed forward and backward in time, a part of this force will act on the test particle at the moment of its acceleration.
The biggest contribution will come from the long-ranged parts of the force:
Every particle in the shell segment $\text d\Omega$ with distance $r$, under the long-ranged force from the test particle, undergoes an acceleration of the same order
and produces an own force of order $\frac1r\cdot\frac1r$, so that the total force applied to the test particle is of order $r^2\text d\Omega\frac1{r^2}=\text d\Omega$ -
 independent of the distance of the shell. Hence, there is a good chance the requirements stated in issue (I1) can be fulfilled.
Assuming the shell behaves as described and one has the special situation of a large acceleration of the test particle,
energy conservation (which, in the sense explained in \cite{WF2}, is valid in Wheeler-Feynman electrodynamics) lets one expect that,
this force has the character of a friction, leading to energy dissipation from the test particle to the shell.
 A crucial difference to usual  friction like Ohm's law is that it can be exerted even on particles with a lot of empty space around them by arbitrarily distant matter. 
In that sense the statistical analysis is different from the usual statistical analysis in which one derives for example  Boltzmann's equation\cite{Spohn2}. 
In Wheeler-Feynman electrodynamics the distant matter does play a role.

As we are going to elaborate, Wheeler and Feynman (\cite{WF1}) investigated this effect quantitatively and found (under simplifying assumptions) that,
independently of the physical properties of the environment, its reactive force near the source has the value $\frac12\left({ F_-}-{ F_+}\right)$
 for ${ F_\pm}$ denoting the Liénard-Wiechert expressions of the test particle.
Its evaluation at the position of the test particle gives rise to the  radiation damping law \eqref{LD}; see \cite{Dirac38}.
Along the way, their result shows that a second test particle, besides its own radiation damping term, 
would feel the total force $\frac12\left({ F_-}+{ F_+}\right)+\frac12\left({ F_-}-{ F_+}\right)= F_-$, 
which justifies the usual effective description of interacting particles via the full retarded Liénard-Wiechert fields, often referred to as causality condition.
Thus, their calculations are a first step towards a possible understanding of irreversible radiation as a thermodynamic phenomenon. 
We remark that, having identified radiation damping as dissipation of energy to the environment, one is led to look for a corresponding fluctuation term, which has not been found yet.

\section{Calculation of radiation damping}\label{sec:radiation-damping}
We think it is helpful to repeat the crucial steps of Wheeler and Feynman (\cite{WF1}) in deriving the radiation damping here.
For simplicity, we focus on their non-relativistic calculations which they call ``Derivation I and II''.
Like Wheeler and Feynman, we refer to the test particle as source, and we model the 
$N-1$ particles surrounding the test particle by a homogenous and not too dense particle distribution, which we referred to as shell in the previous section.
The source has mass $m_0$ and charge $e_0$, and the ``special situation'' is modeled by equipping the source with a periodic acceleration ${ a_0} e^{i\omega t}$
while the surrounding particles are at rest or moving slowly with respect to the source. 
The source and the surrounding particles then interact via the long-ranged part of their Liénard-Wiechert forces and only with the electric components as their velocity is approximately zero
-- the short-ranged part will be neglected.
Within this framework, we shall look for an approximate Wheeler-Feynman solution. 
More specifically, we compute the accelerations of all particles at times close to the instant of the initial acceleration of the source and which are due to the reaction of the initial acceleration of
the source.
%
This approximation is justified if the particles surrounding the source are moving slowly and equilibrium-like whereas the source acceleration is larger than typical.

We emphasize again that we speak of the direct forces in the language of fields which is convenient for the computation.
We put the coordinate origin to the initial location of the source.
According to \eqref{LW3d} (with $v=0$),  the (full) advanced and retarded field a surrounding particle with position ${ x}$ recieves from the source at time $t$ is
$\frac{e_0{ n}\times({ n}\times{ a})}{c^2\|x\|}$, thus has modulus $\frac{e_0\|a\|\sin\angle(a,x)}{\|x\|c^2}$ and is oriented within the $(a,x)$ plane.
Only the component parallel to ${ a}$, which is
\begin{equation}\label{avret}
 { F_\pm}(t,{x}) = -\frac{e_0{ a}\left(t\pm\frac {\|x\|}c\right)\sin^2\angle({a_0},{x})}{\|x\|c^2} \,,
\end{equation}
gives rise to a reaction of the \emph{surrounding particles} and will be referred to as the advanced/retarded field of the source.
Now, $\frac12({ F_+}+{ F_-})$, the elementary field produced by the source, need not be the actual field going from the source to a surrounding particle:
For example, also the advanced fields other surrounding particles produce in response to the half retarded source field are superposed on it.
Therefore, we make the more general ansatz that the total field mediating between the source and the surrounding particles is of the form
\begin{equation}\label{eq:outgoing}
{ F} = \alpha { F_-}+\beta { F_+}\,.
\end{equation}
First, we will consider the special case $\alpha=1$, $\beta=0$; of course, such a choice can only give rise to a Wheeler-Feynman solution if the always existing half advanced source field is later found to be compensated by fields of surrounding particles.

%
The source will act on each surrounding particle by accelerating it, and each surrounding particle interacts with every other surrounding particle. Hence, the effective force of 
the source on one surrounding particle becomes a superposition of many forces  too difficult to describe in a detailed way.
One way out (which needs further scrutiny in a more detailed treatment) is to use  the result from the Drude-Lorentz model which in the language of fields states:
Macroscopic waves of angular frequency $\omega$ in a gas of almost free charges
do not move with speed $c$ but $\frac cn$ with refractive index $n$ given by
\begin{equation}\label{index}
 n=1-\frac{2\pi Ne^2}{m\omega^2},
\end{equation}
where in this section $N$ denotes the number density of the surrounding particles.
In addition to the dispersive behavior it will be convenient to include absorption in our model as well which can be tuned by the absorption coefficient $\gamma$.

In our computation, $ F$ will be treated as ''macroscopic`` in this sense that it is composed out of all the fields emitted by all the particles -- if it were ''microscopic``,
it could not be fully retarded, and the choice $\alpha=1$, $\beta=0$ would not make sense as we mentioned above --
whereas, for the surrounding particles all the ``microscopic'' fields will be identified separately and summed up by hand.

With these model specifications, the field acting on a surrounding  particle at time $t$ is
\begin{equation*}
 { F}(t,{x}) = -\frac{e_0{ a_0}e^{i\omega\left(t-\frac {n\|x\|}c\right) - \gamma\frac{n\|x\|}c}\sin^2\angle({a_0},{x})}{\|x\|c^2} \,.
\end{equation*}
According to \eqref{LW3d}, its corresponding acceleration ${ b} = \frac em { F}$ gives rise to an elementary half advanced and retarded field propagating with $c$.
In order to determine the value of its advanced part ${ F_x}$ (which is part of the ''outgoing`` field ${ F}$ in \eqref{eq:outgoing}) at a position ${ z}$, we need the distance between ${ x}$ and ${ z}$.
If $\|z\|$ is much smaller than $\|x\|$ (which, due to the infinite size of the cloud of surrounding particles, is true for fixed $z$ and ''typical'' values of $x$), then $\|x-z\|\approx \|x\|-\|z\|\cos\angle({z},{x})$, so
\begin{equation*}
 { F_x}(t,z) \approx -\frac{e{ b}\left(t+\frac {\|x\|-\|z\|\cos\angle({z},{x})}c\right)}{2\|x\|c^2}
 \approx \frac{e^2e_0{ a_0}e^{i\omega\left(t+\frac {(1-n)\|x\|}c-\frac {\|z\|\cos\angle({z},{x})}c\right) - \gamma\frac{n\|x\|}c}\sin^2\angle({a_0},{x})}{2m\|x\|^2c^4} \,.
\end{equation*}
Now we can determine the total advanced field emitted by the surrounding particles
\begin{equation*}
 { F_a}(t,z) = N\int { F_x}(t,z)\text d{ x}
\end{equation*}
at a given point $z$ near the source. In spherical coordinates with $\text d{ x}=r^2\text dr\text d\Omega$, we get
\begin{align}\label{eq:integral}
{ F_a}(t,z) =& \frac{Ne^2e_0{ a_0}e^{i\omega t}}{2mc^4} \int_0^\infty e^{i\omega\left(\frac {(1-n)r}c\right) - \gamma\frac{nr}c} \text dr
\int_{\Omega} e^{-i\omega\frac {z\cos\angle({z},{x})}c}\sin^2\angle({a_0},{x}) \text d\Omega \\
=& \frac{Ne^2e_0{ a_0}e^{i\omega t}}{2mc^4}\frac{c}{\gamma n-i\omega(1-n)} \int_{\Omega} e^{-i\omega\frac {z\cos\angle({z},{x})}c}\sin^2\angle({a_0},{x}) \text d\Omega \,.\nonumber
\end{align}
Letting $\gamma$ go to zero and using \eqref{index} for the refractive index n,
\begin{equation}\label{absfeld}
 { F_a}(t,z) = \frac{i\omega e_0{ a_0}e^{i\omega t}}{4\pi c^3} \int_{\Omega} e^{-i\omega\frac {\|z\|\cos\angle({z},{x})}c}\sin^2\angle({a_0},{x}) \text d\Omega
\end{equation}
remains.

We observe that $n$ makes the integral in \eqref{eq:integral} finite and $\gamma$ makes it definite.
Note that the absorption coefficient $\gamma$ does not arise from the calculations within the Drude-Lorentz model as the unphysical situation of an infinite plane wave hitting an infinite medium
is considered, and already there, infinite integrals like in \eqref{eq:integral} for $\gamma=0$ are not treated with sufficient care (see \cite{FeynmanLectures}, ch. 31).
We expect that a more careful analysis will yield the absroption coefficient $\gamma$ as naturally as the refractive index $n$.
It would therefore be rash to conclude just from the appearance the divergence in \eqref{eq:integral} for $\gamma=0$
that the whole universe is part of the mechanism and has to have properties of an absorbing medium and to draw cosmological conclusions in the way, e.g., \cite{Hogarth, PCWDavies, HoyleNarlikar} do.

The formula for the complex refractive index that Wheeler and Feynman use in \cite[Derivation II]{WF1}  makes things formally look cleaner, however we have a good reason to avoid it:
Whereas the Drude-Lorentz model can deal with a classical low-density electron gas and therefore, there is no principle obstacle to making it rigorous in the Wheeler-Feynman framework,
the derivation of the latter formula relies on the polarization of atoms in the medium which cannot be described with classical Wheeler-Feynman electrodyanmics.

Back to the calculation -- for the remaining integral, we use $\cos\angle({z},{x})$ and the angle between the $({ z}, { a_0})$ and $({ z}, { x})$ plane ($\varphi$) as coordinates.
Then $\frac{\text d\Omega}{4\pi} = \frac12 \text d\cos\angle({z},{x}) \frac{\text d\varphi}{2\pi}$ and
\begin{equation}\label{int}
\int_{\Omega} e^{-i\omega\frac {\|z\|\cos\angle({z},{x})}c}\sin^2\angle({a_0},{x}) \frac{\text d\Omega}{4\pi} =
\frac12 \int_{-1}^1 e^{-i\omega\frac {\|z\|\cos\angle({z},{x})}c} \left(\int_0^{2\pi}\sin^2\angle({a_0},{x}) \frac{\text d\varphi}{2\pi}\right) \text d\cos\angle({z},{x}) 
\end{equation}
holds.
To compute the inner integral, the sine has to be expressed in terms of $\varphi$ and $\cos\angle({z},{x})$ (and constants like $\angle({a_0}, {z})$).
To do so, we use unit vectors in the ${ z}$, ${ a_0}$ and ${ x}$ directions in suitable coordinates.
If we choose ${ \hat z}=\begin{pmatrix}1\\0\\0\end{pmatrix}$ and ${ \hat a_0}=\begin{pmatrix}\cos\angle({z},{a_0}) \\ \sin\angle({z},{a_0}) \\ 0\end{pmatrix}$,
then ${ v}=\begin{pmatrix}0\\ \cos\varphi \\ \sin\varphi \end{pmatrix}$ is a unit vector in the $({ z},{ a_0})$ plane perpendicular to ${ z}$,
so $${\hat x} = \cos\angle({z},{x}) {\hat z} + \sin\angle({z},{x}) { v} = \begin{pmatrix}\cos\angle({z},{x}) \\ \sin\angle({z},{x}) \cos\varphi \\ \sin\angle({z},{x}) \sin\varphi \end{pmatrix}\,.$$
Consequently,
\begin{align*}
 \sin^2\angle({a_0},{x}) = 1-\langle{\hat a_0},{\hat x}\rangle^2
\end{align*}
and
\begin{align*}
 \int_0^{2\pi}\sin^2\angle({a_0},{x}) \frac{\text d\varphi}{2\pi} =& 1-\cos^2\angle({z}, {a})\cos^2\angle({z}, {x}) - \frac12 \sin^2\angle({z}, {a})\sin^2\angle({z}, {x}) \\
=& 1-\frac12\sin^2\angle({z}, {a}) + \left(\frac32\sin^2\angle({z}, {a}) - 1\right)\cos^2\angle({z}, {x}) \,.
\end{align*}
Substituting into \eqref{int}, 
\begin{equation}\label{domega}
 \begin{split}
  \int_{\Omega} e^{-i\omega\frac {\|z\|\cos\angle({z},{x})}c}\sin^2\angle({a_0},{x}) \frac{\text d\Omega}{4\pi} 
=& \frac12 \int_{-1}^1 e^{-i\omega\frac {\|z\|s}c} \left( 1-\frac12\sin^2\angle({z}, {a}) + \left(\frac32\sin^2\angle({z}, {a}) - 1\right)s^2 \right) \text ds \\
=& \sin^2\angle({z}, {a}) \frac{e^{iu}-e^{-iu}}{2iu} +  \left(3 \sin^2\angle({z}, {a}) -2\right) \left(\frac{\cos u}{u^2}-\frac{\sin u}{u^3}\right)
\end{split}
\end{equation}
with $u=\frac{\omega \|z\|}c$, so that -- for sufficiently large distances $\|z\|$ - \eqref{absfeld} and \eqref{avret} lead to
\begin{equation}\label{absfeld2}
 { F_a}(t,z) = \frac{ e_0{ a_0}e^{i\omega t}\sin^2\angle({z}, {a})}{2zc^2}\left(e^{\frac{i\omega z}{c}}-e^{-\frac{i\omega z}{c}}\right) = \frac12\left({ F_-}(t,z) - { F_+}(t,z)\right) \,.
\end{equation}
Moreover, applying  de l'Hospital  to \eqref{domega} for $u\to0$ and again substituting into \eqref{absfeld}, one gets
\begin{equation}\label{damping}
 { F_a}(t,0) = \frac{2i\omega e_0{ a_0}e^{i\omega t}}{3c^3} = \frac{2e_0}{3c^3}{ \dot a}(t) \,,
\end{equation}
which coincides with the non-relativistic limit of the radiation damping force \eqref{LD} on the source.

To summarize, we have found a bona fide Wheeler-Feynman solution: 
The surrounding particles move according to the total (purely retarded) field ${ F}$ acting on them, and the decomposition
\begin{equation*}
 { F} = { F_-} = \frac12({ F_-}+{ F_+}) + \frac12({ F_-}-{ F_+})= \frac12({ F_-}+{ F_+}) +{ F_a}
\end{equation*}
shows that the assumed ${ F}$ is precisely the field produced by the source and the surrounding particles with the given accelerations together.
The damping force \eqref{damping} contributes only a small correction to the primary source acceleration and thus to $\frac12({ F_-}+{ F_+})$.

At first sight, one might wonder whether the retarded fields the surrounding particles  produces in response to the retarded field $ F$ have a similar effect on the source as $ F_a$
some time after its acceleration.
This is not the case because the fields moving further forward in time reach the source at completely different instants and cannot be expected to have a visible common effect.
It is beyond our present scope to say anything about the further time evolution, but we expect that, after the external force on the source stops and it has been sufficiently damped,
the motion of the surrounding particles quickly return to equilibrium and the (yet unknown) fluctuation effects come into play.

Note that the advanced fields of the surrounding particles are not cancelled by the same reasoning as for of the source:
It is crucial for the calculation that the source acceleration is assumed to be so large that the fields produced by equilibrium fluctuations of the surrounding particles are negligible in comparison
to the response to the source acceleration.
The acceleration of a surrounding charge, in turn, is expected to deviate little from an equilibrium value.

So much about $\alpha=1\,,\beta=0$.
In order to repeat the calculation with $\alpha=0\,,\beta=1$, one has to compute retarded instead of advanced responses, which only amounts to changing some signs.
Our calculations show that the response of the surrounding particles is linear in the field strength of ${ F}$ (as long as its direction remains the same), so that, for general $\alpha$ and $\beta$,
we get
\begin{equation*}
 { F_a} = \frac\alpha2({ F_-}-{ F_+}) + \frac\beta2({ F_+}-{ F_-}) = \frac{\alpha-\beta}2 { F_-} + \frac{\beta-\alpha}2 { F_+} \,.
\end{equation*}
Such an $ F_a$ can only belong to an approximate Wheeler-Feynman solution if the total field decomposes into
\begin{equation*}
 { F} = \alpha { F_-}+\beta { F_+} = \frac12({ F_-}+{ F_+}) + { F_a} \,,
\end{equation*}
leading to the requirement $\alpha+\beta=1$. Indeed, all environmental fields
\begin{equation*}
 { F_a} = \left(\alpha-\frac12\right){ F_-} + \left(\frac12-\alpha\right){ F_+}
\end{equation*}
with corresponding particle accelerations are approximate Wheeler-Feynman solutions.
For example, the one with $\alpha=0$ is the time reversal of the one with $\alpha=1$, which means that there is a synchronized motion in the environment before the source acceleration that
leads to ``inverse radiation damping'' at the moment of the source acceleration.

\section{Time arrow, irreversibility}
 Wheeler-Feynman electrodynamics is time symmetric, i.e., the time reversed solution is also a solution.
However, in non-equilibrium situations the latter may look extremely conspiratorial. For this the following analogy is useful again:
Classical mechanics can describe a glass breaking into pieces, it  can also describe the pieces forming the glass by time reversal.
The ``glass breaking into pieces'' is typical for our universe and is accompanied by an increase of entropy, which defines a thermodynamical arrow of time. 
 It can be reduced to and in this sense explained by a particular low-entropy state of matter  in which our universe started.
The idea of Wheeler and Feynman is to hold the very same low entropy state responsible for the observed retarded radiation and radiation friction, i.e., 
the fact that only the ``$\alpha=1$ solution`` is observed. 
 
%
The global low entropy state of matter can be thought of as an inhomogenous matter distribution in which atypical accelerations in ''subsystems'' can be created.
Either there are local ``equilibrium islands'' within this distribution, or it is sufficiently homogenous on a larger scale relevant for the preceding calculation --
in both cases, as in classical mechanics, a ``heat bath`` is always present. Therefore,
 atypical accelerations can relax by interaction with the surrounding matter ``heating up their motion'', and can effectively be described by retarded radiation and global  increase of entropy.
However, to really get a  grip on the latter we must, as we already said, understand better how entropy is defined in a Wheeler-Feynman universe.

  In particular, the statistical analysis of the theory, e.g., in terms of phase space measure, is lacking
in order to ascertain the consistency of global non-equilibrium and sufficient homogeneity properties in the way described.
Whether all the future environment of a subsystem must plays a role for radiation damping
(which might give rise to the speculation that radiation damping is already an account of the heat death of the universe)
or only local parts of it  also remains to be scrutinized further.

We would also like to remark in this respect, that it would be most desirable to have a similar 
analysis as in section \ref{sec:radiation-damping} for two test particles accelerating each other and proving that 
the effect of the environment yields that the analogous effective description as of one test particle.

\section{The ``absorber condition''}

In Wheeler-Feynman electrodynamics no radiation in the sense of ``escaping fields'' exists.
As explained, what we referred to as radiation is actually a thermodynamic phenomenon in many-particle systems -- in particular,
there would be no radiation effects in a Wheeler-Feynman universe consisting only of a few particles.
As we have also emphasized, Wheeler and Feynman's computation of radiation damping is completely independent of the precise arrangement of the surrounding particles as well as their physical properties
like their particular masses or charges.
This fact might suggest that radiation damping can actually be inferred from a more abstract argument.
To establish this goal and inspired by Tetrode's ideas (\cite{Tetrode}) Wheeler and Feynman argued in \cite[Derivation IV]{WF1} that the crucial role of the surrounding particles is to \textit{absorb}
the field of the test particle, which can motivated as follows:


For a system of $N$ particles ($N$ being large) with particle $i$ being the test particle, a summary of the picture elaborated in Section \ref{sec:introduction} is that motion of the $j\neq i$ other
particles is equilibrium-like before before particle $i$ is accelerated.  In particular, this means
\begin{equation}\label{ggw}
 \sum_{j\neq i}{ F_{j,-}}\approx0\,.
\end{equation}
 The
 absorber response to the atypical acceleration of the test particle $i$ is given by \eqref{absfeld2}, i.e.,
\begin{equation*}\label{absr}
 \frac12\sum_{j\neq i}{ F_{j,+}} \approx \frac12 ({ F_{i,-}}-{ F_{i,+}}) \,.
\end{equation*}
Subtracting twice the second equation from the first one, one obtains
\begin{equation}\label{absbed-}
 \sum_j ({ F_{j,-}}-{ F_{j,+}}) \approx 0 \,.
\end{equation}
One then observes that using this together with \eqref{ggw}
 immediately recovers  \eqref{LD} in the form
\begin{equation}\label{eq:friction}
  m\ddot { x}(t) \approx \frac{2e^2}{3c^3}\dddot { x}(t)\,.
\end{equation}

This looks like a promising starting point  to abstract the whole  analysis of radiation damping based on the condition \eqref{absbed-}. 
%
%
For this purpose Wheeler and Feynman argue in \cite[Derivation IV]{WF1} how one can arrive at \eqref{absbed-}  without any particular model for the absorber, namely  only by assuming 
\begin{equation}\label{absbed+}
 \sum_j({ F_{j,+}}+{ F_{j,-}}) = 0 \text{ outside the absorber at all times.}
\end{equation}
This condition says that a test particle outside of the absorber (or ``far away from the center'' if the extent of the absorber is infinite) feels no force and is therefore often referred to as
``absorber condition''.
To recover \eqref{absbed-} from \eqref{absbed+} Wheeler and Feynman argue further that, since $\sum { F_{j,-}}$ represents an incoming and $\sum { F_{j,+}}$ an outgoing wave, \eqref{absbed+}
implies that already each of them is zero individually.
Therefore \eqref{absbed-} has to hold ``outside'' for all times and as \eqref{absbed-} is a solution to the homogeneous Maxwell equations, it must also be fulfilled everywhere.

One could now be inclined to think that in order to recover radiation damping in a Wheeler-Feynman universe,
\eqref{absbed+} must be fulfilled a priori and in addition to the fundamental equation \eqref{Gleichung}. In this respect we remark:

\begin{enumerate}
  \item As we have explained in Sections \ref{sec:introduction} and \ref{sec:radiation-damping},
radiation damping can already be inferred from \eqref{Gleichung} by a statistical reasoning alone -- in particular, without any extra condition like \eqref{absbed+} or \eqref{absbed-}.

  \item The reasoning that led from \eqref{absbed+} to \eqref{absbed-} is based on exact equalities.
After looking at the elaborated statistical reasoning and, in particular because it is not clear how all fields arising from fluctuations and inhomogeneities near the boundary of the the cloud
of $j\neq i$ particles could be cancelled, one has to take the possibility into account that
the absorber condition \eqref{absbed+} might be satisfied only approximately. In such a case, it seems to be far from clear how one can get such a good estimate in \eqref{absbed-} that the corresponding disturbance in the radiation damping effect, which is small anyway, is smaller than the effect itself.
  
  \item In contrast, nothing should prevent us from dealing with disturbances caused by fluctuations in the statistical reasoning given in in Sections \ref{sec:introduction}
and \ref{sec:radiation-damping}; there, we do not rely on a notion of smallness which has to be controlled over the whole time evolution of the universe.
  

 \end{enumerate}


The status of the absorber condition \eqref{absbed+} is therefore dubious, but presumably also not relevant to explain radiation damping.
For that reason we think that the sole focus on it, as one finds it in textbooks on Wheeler-Feynman electrodynamics and its mentioned cosmological considerations, is off target.

\vspace{1cm}
{\bf Acknowledgements: }We would like to thank Sheldon Goldstein, Michael Kiessling, Steve Lyle, and  Herbert Spohn for stimulating discussions.

\vspace{0.5cm}

\bibliographystyle{plain}

\end{document}